\documentclass{aa}  
\usepackage{graphicx,natbib}
\usepackage{txfonts}
\bibpunct[, ]{(}{)}{,}{a}{}{,}
\begin{document}
   \title{A rapid and dramatic outburst in Blazar 3C\,454.3 during May 2005}

   \subtitle{Optical and infrared observations with REM and AIT}

   \author{L. Fuhrmann\inst{1,2}, A. Cucchiara\inst{3}, N. Marchili\inst{1}, G. Tosti\inst{1}, G. Nucciarelli\inst{1}, 
           S. Ciprini\inst{4,1}, E. Molinari\inst{5}, G. Chincarini\inst{5,6}, F. M. Zerbi\inst{5},  S. Covino\inst{5}, 
	   E. Pian\inst{7}, E. Meurs\inst{8}, V. Testa\inst{9}, F. Vitali\inst{9}, L. A. Antonelli\inst{9}, P. Conconi\inst{5}, 
	   G. Cutispoto\inst{10},\\ G. Malaspina\inst{5}, L. Nicastro\inst{11}, E. Palazzi\inst{12}, P. Ward\inst{8}\\
} 

   \offprints{L. Fuhrmann,\\ e-mail: lars.fuhrmann@fisica.unipg.it}

   \institute{Dipartimento di Fisica e Osservatorio Astronomico, Universit\`a di Perugia, Via A. Pascoli, 06123 Perugia, Italy
     \and     INAF, Osservatorio Astronomico di Torino, Via Osservatorio 20, 10025 Pino Torinese (TO), Italy
     \and     Department of Astronomy and Astrophysics, Pennsylvania State University, 525 Davey Laboratory, PA 16802, USA 
     \and     Tuorla Astronomical Observatory, University of Turku, V\"{a}is\"{a}l\"{a}ntie 20, 21500 Piikki\"{o}, Finland
     \and     INAF, Osservatorio Astronomico di Brera, Via E. Bianchi 46, 23807 Merate (Lc), Italy
     \and     Universit\'a degli studi di Milano-Bicocca, Dipartimento di Fisica, Piazza delle Scienza 3, 20126 Milan, Italy
     \and     INAF, Osservatorio Astronomico di Trieste, Via G. B. Tiepolo 11, 34131 Trieste, Italy
     \and     Dunsink Observatory, Castleknock Dublin 15, Ireland
     \and     INAF, Osservatorio Astronomico di Roma, Via Frascati 33, 00040 Monteporzio Catone, Italy
     \and     INAF, Osservatorio Astrofisico di Catania, Via S. Sofia 78, 95123 Catania, Italy
     \and     IASF/CNR, Sezione di Palermo, Via Ugo La Malfa 153, 90146 Palermo, Italy
     \and     IASF/CNR, Sezione di Bologna, Via Gobetti 101, 40129 Bologna, Italy
}
   
   \date{September 2005}

   \abstract{
The flat-spectrum radio quasar 3C\,454.3 is well known to be a highly active and variable 
source with outbursts occurring across the whole electromagnetic spectrum over the last decades. 
In spring 2005, 3C\,454.3 has been reported to exhibit a strong optical outburst which 
subsequently triggered multi-frequency observations of the source covering the radio up to $\gamma$-ray bands. Here, 
we present first results of our near-IR/optical (V, R, I, H band) photometry performed between May 11 and August 5, 
2005 with the Rapid Eye Mount (REM) at La Silla in Chile and the Automatic Imaging Telescope (AIT) of the Perugia 
University Observatory. 3C\,454.3 was observed during an exceptional and historical high state with a subsequent 
decrease in brightness over our 86\,days observing period. The continuum spectral behaviour during the flaring and 
declining phase suggests a synchrotron peak below the near-IR band as well as a geometrical origin of the variations 
e.g. due to changes in the direction of forward beaming. 
   
   \keywords{galaxies: active -- galaxies: blazars: general -- galaxies: blazars: individual: 3C\,454.3 -- galaxies: jets 
     -- galaxies: quasars: general 
               }
   }
   \authorrunning{L. Fuhrmann et al.}
   \maketitle
%
%________________________________________________________________

\section{Introduction}

The blazar sub-class of radio-loud Active Galactic Nuclei (AGNs) consists of
flat-spectrum radio quasars (FSRQ) and BL Lacs which exhibit rather complex
and outstanding characteristics such as extreme variability at all
wavelengths, polarisation, strong $\gamma$-ray emission, often high
superluminal motion and brightness temperatures exceeding the inverse Compton
limit \citep[e.g.][]{1999APh....11..159U}. The overall spectral energy distribution 
(SED) of blazars is characterised by two broad components. The first one peaks in the IR-
soft-X-ray band and is due to synchrotron emission, while the second component 
at higher energies is most likely produced by inverse Compton emission. The
mechanism for the rapid and high amplitude variability often observed in blazars 
is still not well understood. Different theoretical models are discussed which 
include e.g. shock-in-jets \citep[e.g.][]{1996ASPC..100...45M}, colliding relativistic plasma 
shells \citep[e.g.][]{2004A&A...421..877G} as well as changes in the direction of forward 
beaming \citep[e.g.][]{1999A&A...347...30V}.       

The flat-spectrum radio quasar 3C\,454.3 ($z=0.859$) is a typical example
of a highly variable blazar source. It is one of the brightest extragalactic 
radio sources on the sky and multi-epoch VLBI studies revealed a compact, 
superluminal ($\beta_{app}\lesssim$\,8c) source showing a bent, one-sided 
core-jet structure on mas scales \citep[e.g.][]{1995PNAS...9211377K,1998ASPC..144...75P,
2004evn..conf....7P}. Strong flux density variability in the 
radio regime was observed during long-term monitoring campaigns carried out 
over the last decades \citep[e.g.][and ref. therein]{1997AIPC..410.1423A,
1998A&AS..132..305T,2004A&A...419..485C}. While its low-frequency (decimeter) 
variability was partly attributed to interstellar scintillation \citep[e.g.][]
{1984AJ.....89.1784A}, the source often shows strong synchrotron outbursts on time scales 
of years in the cm- to mm-regime. These outbursts at different frequencies 
appear correlated often with increasing variability amplitude towards higher 
frequencies \citep[e.g.][]{1994A&A...289..673T}. Here, a strong and consistent 
periodicity of about six years was found in the overall flaring behaviour of the source  
\citep{2004A&A...419..485C}. In addition, new VLBI component ejections were
suggested to occur contemporaneously with radio outbursts and enhanced levels
of $\gamma$-ray flux \citep{1995PNAS...9211377K,2004evn..conf....7P}.
At high energies, 3C\,454.3 is a prominent example of a strong and variable 
source detected by e.g. Einstein, ROSAT, COMPTEL, EGRET and OSSE with a spectral 
maximum occurring at MeV energies \citep[e.g.][]{1995A&A...295..330B,1996ApJS..105..331L}.    
%______________________________________________________________
%
   \begin{figure}
   \centering
   \vspace{-0.3cm}
   \includegraphics[width=6.2cm,angle=-90]{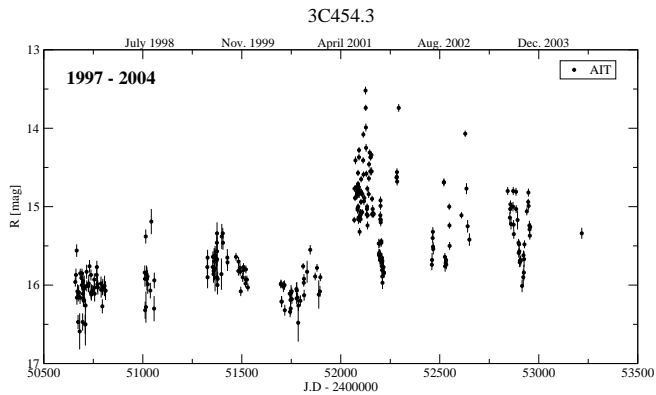}
   \vspace{-0.3cm}    
   \caption{Long-term light curve of 3C\,454.3 obtained with the AIT telescope during a blazar 
	monitoring project between 1997 and 2004. The source shows a very active behaviour with 
	different variability amplitudes and time scales. The most prominent outburst occurred in 
	August/September 2001. R band data are shown as this band is most commonly observed.}
       \label{Fig_Longterm}
   \end{figure}
%______________________________________________________________ 

In the optical band, the source was early identified as an optically violently
variable. Historical B band data go back to 1900 \citep{1968PASP...80..339A} 
and since then several outbursts have been reported over the decades (see also Fig. 
\ref{Fig_Longterm}). High optical polarisation ($P_{opt}=0-16$\,\%) and 
polarisation angle changes were observed \citep[see][]{1980ARA&A..18..321A}. 
Several attempts were made to find a correlation between the radio and optical events 
in 3C\,454.3. \citet{1994A&A...289..673T} reported a possible correlation for
mm-radio and optical data obtained between 1980 and 1993 confirming the previous 
findings of  \citet{1976AJ.....81..489P} and \citet{1982PhDT.........6B}. The 
latter authors suggested a correlation for the optical variations leading the 
radio events with a time lag of about 1.2\,yr with a tendency of considerably 
shorter time scales for the optical variability. In addition, intra-night
variability of 3C\,454.3 at optical bands was detected on several occasions
\citep[$\sim$\,0.3\,mag/hour;][]{1971AJ.....76...25A}. During a monitoring campaign of
$\gamma$-ray loud blazars, 3C\,454.3 was found to exhibit rapid R band
variations of 0.3\,mag in less than 3 hours without comparable behaviour 
in other bands \citep{1997A&AS..121..119V,1998A&AS..127..445R}.

In spring 2005 a new, exceptional strong outburst in 3C\,454.3 has been reported
\citep{vsnet-alert8383,vsnet-alert8405} which triggered subsequent, quasi-simultaneous broad band
observations of the source at radio, IR and optical bands as well as at higher 
energies \citep[X-ray/$\gamma$-ray,][]{2005ATel..497....1F}. Here, we present first results 
of our multi-band photometry (V, R, I, H bands) performed during the new outburst phase 
between May and August 2005 probably showing the most rapid variations (and historical brightness) 
observed in this source so far. Comprehensive papers in preparation will present and 
analyse this data in the framework of a broad band spectral energy approach using 
the quasi-simultaneous observations at cm-to mm-wavelengths as well as data obtained 
at higher energies with the SWIFT and INTEGRAL satellites \citep{Giommi,Pian,Fuhrmann}. 

\section{Observations and data reduction}
The photometric optical/IR monitoring of 3C\,454.3 started on May 11, 2005
in the V, R, I, and H band. The last data presented here were obtained on
August 5, 2005 and thus our observations cover a total time period of 86\,days 
(J.D. 2453502.4--2453588.0). The observations were carried out with two
instruments: (i) the Newtonian f/5, 0.4\,m, Automatic Imaging Telescope (AIT)
of the Perugia University Observatory, Italy and (ii) the Rapid Eye Mount 
\citep[REM,][]{2004SPIE.5492.1590Z}, a robotic telescope located at the ESO Cerro La 
Silla observatory in Chile. The AIT is based on an equatorially 
mounted 0.4-m Newtonian reflector having a 0.15-m refractor solidly joined 
to it. AIT is a robotic telescope equipped with a 192$\times$165 pixels CCD
array, thermo-electrically cooled with Peltier elements and Johnson-Cousins
BVRI filters are utilised for photometry \citep{1996PASP..108..706T}. The 
REM telescope has a Ritchey-Chretien configuration with a 60\,cm f/2.2 primary 
and an overall f/8 focal ratio in a fast moving alt-azimuth mount providing two
stable Nasmyth focal stations. At one of the two foci the telescope simultaneously 
feeds, by means of a dichroic, two cameras: REMIR for the NIR \citep[see][]
{2004SPIE.5492.1602C}, and ROSS \citep[see][]{2004SPIE.5492..689T} for the 
optical. Both cameras have a field of view of 10$\times$10 arcmin and imaging 
capabilities with the usual NIR (z', J, H and K) and optical broad band 
(V, R, I) filters. Moreover, via an Amici prism, low resolution slitless 
spectroscopy is also possible. 

At the AIT telescope images were taken at R and I bands during a total of 30
nights, whereas V, R, I, H band REM observations were performed during 23
nights of the 86\,days observing period (see Table \ref{summary}). The data 
were usually taken as a sequence of two or more frames per band and
night. Each single H band observation with REMIR was performed with a 
dithering sequence of five images shifted by a few arcsec. These 
images are automatically elaborated using the jitter script of the ECLIPSE
package. The script aligns the images and co-adds all the frames to obtain 
one average image for each sequence.    

After dark-, bias- and flat-field correction, each image of both telescopes was
reduced using standard procedures. A combination of the DAOPHOT \citep{1988PASP...99..191F}
and Sextractor packages \citep{1996A&AS..117..393B} was used to perform standard aperture 
photometry. The instrumental magnitudes were then further processed to obtain 
standard magnitudes of 3C\,454.3 and the relevant errors through differential 
photometry. Here, an average zero-point correction of each optical frame was applied 
using the tabulated magnitudes of 5--8 standard reference stars in the comparison 
star sequence reported by \citet*{1998PASP..110..105F} and \citet{1998A&AS..130..495R}.
The calibration of the NIR data was done using the comparison star sequence published by 
\citet{2001AJ....122.2055G}. Errors on the calibrated magnitudes of 3C\,454.3 were 
calculated taking into account the statistical error of the reduction process 
(aperture photometry) plus the rms of the variations as seen in the reference 
star zero-points. For the subsequent spectral analysis the data sets 
were corrected for galactic extinction according to \citet{1998ApJ...500..525S}.

The calibrated light curves of the two telescopes were then combined in the R and I band, 
respectively. Here, a comparison among the data acquired with the two different telescopes in 
the same night did not show significant instrumental offsets. These offsets have been 
computed, for each band, from data acquired contemporaneously by the two telescopes 
(simultaneous observations occured only in 5 nights). All values were found to lie within 
the individual magnitude errors and thus were negligible. The final light curves of all bands 
are presented in Fig. \ref{Fig_all_bands} while Table \ref{summary} gives a short summary of 
the observations and results for each band by means of average magnitudes and flux densities 
$F_{\nu}$.
%_____________________________________________________________
\begin{table}
\begin{center}
\caption{A short summary of the observations for each band performed between May and
  August 2005. Here, $T_{obs}$ denotes the observing duration and $N_{data}$
  the total number of data points obtained over this time interval.}
\vspace{-0.5cm}
\begin{tabular}{c|ccccccc}
\hline
\hline
     & $T_{obs}$ & $N_{data}$ & $\langle mag\rangle$ & $\Delta$\,mag & $\langle F_{\nu}\rangle$ & $\Delta\,F_{\nu}$\\
     & [days]    &            &                      &               & [mJy]                    & [mJy]      \\
\hline
V    & 60        & 48         & 14.16$\pm$0.73 & 2.37 & 13.7$\pm$9.7  & 36.6\\
R    & 86        & 85         & 13.57$\pm$0.63 & 2.61 & 17.5$\pm$10.0 & 47.6\\
I    & 86        & 72         & 12.90$\pm$0.65 & 2.46 & 25.4$\pm$15.4 & 64.6\\
H    & 42        & 167        & 10.18$\pm$0.36 & 1.60 & 92.6$\pm$30.7 & 142.6\\
\hline
\hline
\end{tabular}
\label{summary}
\end{center}
\end{table}
%______________________________________________________________

\section{Results and discussion}
\begin{figure}
   \centering
   \vspace{0.5cm}
   \hspace{0.2cm}
   \includegraphics[width=8.2cm,angle=0]{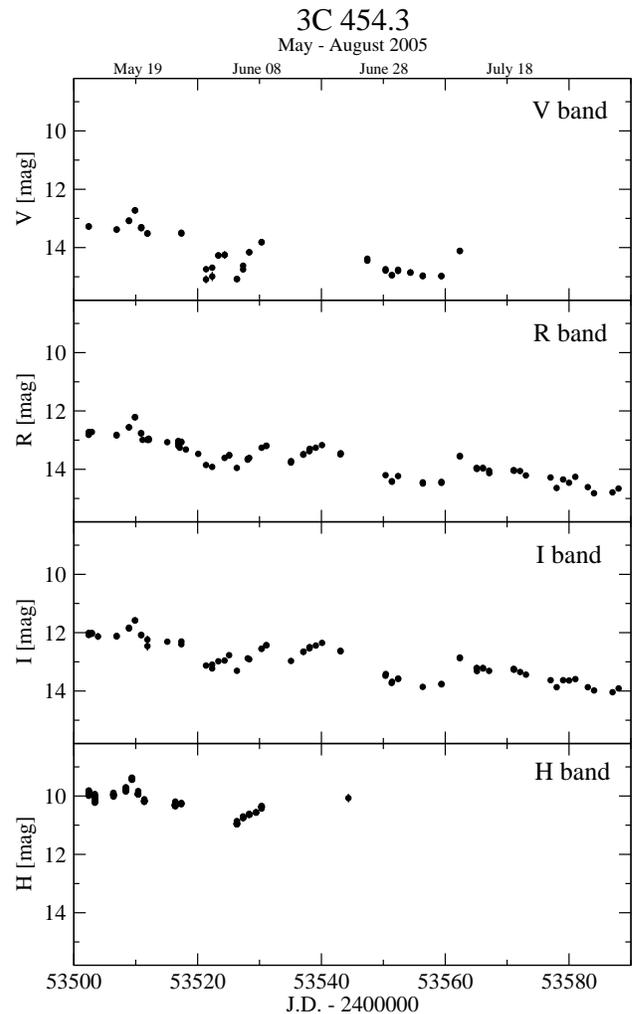}
   \vspace{-0.1cm}    
   \caption{Final optical and near-IR light curves of 3C\,454.3 obtained between
      May and August 2005. The V and H band data cover a significantly shorter
      time interval. All light curves are plotted on the same scale for a better comparison. 
      The photometric errors are often comparable to the symbol size.}
         \label{Fig_all_bands}
   \end{figure}

During the last decade 3C\,454.3 passed through several active states as visible 
in our archival long-term light curve obtained with the AIT telescope between
1997 and 2004. In Fig. \ref{Fig_Longterm}, R band data are shown as this is the band 
most commonly observed. Although rather undersampled, the light curve shows long-term trends 
with faster and stronger outbursts superimposed. The most prominent feature is 
visible around August/September 2001, when the source reached a magnitude of 
$R=13.5$ and subsequently decreased in brightness by $\Delta\,R=1.7$ during a 
period of only about 10\,days. The observations presented here show 3C\,454.3
during a new outburst phase when the source was detected at its historical maximum 
brightness in all bands. All light curves in Fig. \ref{Fig_all_bands} show 3C\,454.3 
in an exceptional bright state at the beginning of May 2005 with a subsequent
long-term decrease over the 86 days observing period as seen in the R and I bands. 
Preliminary data of \citet{vsnet-alert8383,vsnet-alert8405} indicate that the maximum 
brightness (R$\sim$12\,mag) occurred two days before the start of our observations
(May 09, 2005), thus we observed the source during the declining phase of the
outburst. In addition to the long-term decrease in brightness, all bands show several
faster, simultaneous flares superimposed with time scales of often only a few
days. The first occurred around J.D. 2453510 when the source reached its 
maximum in all light curves with V\,=\,12.7, R\,=\,12.2, I\,=\,11.6 and H\,=\,9.4, 
respectively.
The overall highest minimum-to-maximum variations are seen in the R band with 
$\Delta R=2.6$ over a time period of 75 days (see also Table \ref{summary}).   
%______________________________________________________________
%
%______________________________________________________________

\begin{figure}
   \centering
%   \hspace{-0.5cm}
   \includegraphics[width=7.1cm,angle=-90]{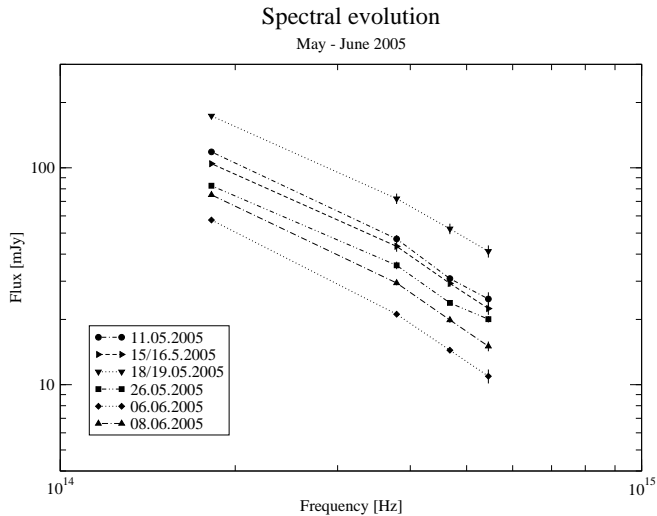}
   \vspace{-0.2cm}
      \caption{Near-IR/optical continuum spectra obtained at six epochs between May 11 and June
      8, 2005. Over this time period no strong, significant spectral changes are evident.}
         \label{Fig_spec_evol}
   \end{figure}

In Fig. \ref{Fig_spec_evol} the continuum spectral evolution over six epochs is 
shown at several levels of intensity. These V to H band spectra were taken during periods 
of temporal overlap of the four bands between May 11 and June 08 (J.D.\,2453502.4--2453530.3). 
Power-law fits to each epoch show no strong and significant spectral changes over this period. 
Such 'achromatic' variability behaviour suggests a geometrical origin of the observed 
variations due to changes in the direction of forward beaming \citep[e.g. a helical/precessing 
jet;][]{1980Natur.287..307B,1999A&A...347...30V} rather than acceleration/cooling mechanisms 
in the jet of 3C\,454.3. However, the steep spectral shape with a mean value of 
$\alpha$\,\,=\,\,1.39\,\,$\pm$\,\,0.07 ($F(\nu)\propto\nu^{-\alpha}$) indicates that 
the maximum of the synchrotron peak was located at lower frequencies than the 
near-IR band during the maximum as well as the declining phase of the outburst. 
Here, the quasi-simultaneous observations obtained at cm- and mm-wavelengths will help 
to study the spectral shape, peak and evolution of the synchrotron part of the
SED of 3C\,454.3 during the recent flare in more detail \citep{Fuhrmann}.

First results of (quasi-) simultaneous broad band observations obtained this year 
indicate that 3C\,454.3 was in a similar active state over the last months. During 
the period of our REM and AIT observations, the SWIFT satellite observed the source 
at three epochs between May 11 and May 19, 2005 as part of an ongoing X-ray investigation 
of a sample of blazar sources. First results are presented by \citet{Giommi} and 
display an exceptional, by a factor of about 10 higher X-ray flux compared to previous
ROSAT observations. The simultaneous REM IR/optical and SWIFT UV/X-ray data show a complex 
variability and SED behaviour which requires a more detailed analysis. 
%______________________________________________________________
%
%______________________________________________________________

\section{Summary and conclusions}
The blazar 3C454.3 went through a dramatic, historical outburst at the beginning of May 2005.  
Our photometric monitoring of the source with the REM and AIT telescopes over a time period of 86\,days 
shows a very bright state at the beginning of May 2005 with a subsequent, long-term decrease in brightness 
with the highest minimum-to-maximum variations of $\Delta R=2.6$. The highest brightness in our V to H band light 
curves was detected during a fast, secondary outburst occurred around May 19, 2005. 
A first analysis of the spectral behaviour of the source across our near-IR/optical bands during six epochs 
between May and June 2005 displays a steep and, in time, nearly constant spectral slope with 
$\langle \alpha\rangle$\,=\,1.39\,$\pm$\,0.07. This suggests (i) a maximum of the synchrotron emission in the SED of 
3C\,454.3 to be below the near-IR band in Fig. \ref{Fig_spec_evol} and (ii) a geometrical origin of the 
variations. Here, changes in the viewing angle of the jet with respect to the line of sight and thus 
temporal changes of the Doppler factor could be invoked. However, a more detailed analysis of the 
synchrotron component will become possible combining our observations with (quasi-) simultaneous 
data obtained at radio- and mm-bands, as will be discussed in a forthcoming paper \citep{Fuhrmann}.

\begin{acknowledgements}

The work presented here was partly supported by the European Institutes belonging to the ENIGMA
collaboration acknowledge EC funding under contract HPRN-CT-2002-00321.                       

\end{acknowledgements}

\end{document}